\documentclass[11pt]{article}

\usepackage[margin=1in]{geometry}
\usepackage{amsmath,amssymb,amsfonts}
\usepackage{graphicx}
\usepackage{booktabs}
\usepackage{url}
\usepackage{enumitem}
\usepackage{bbm}
\usepackage{natbib}
\usepackage[table]{xcolor}
\newcommand{\highlightrow}{\rowcolor{gray!15}}
\usepackage{hyperref}

\title{Random-Forest-Induced Graph Neural Networks for Tabular Learning}

\author{
Haozhe Chen\thanks{Co-first authors and equal contributions.} \\
Utah State University \\
\texttt{haozhe.chen@usu.edu}
\and
Soheila Farokhi\footnotemark[1] \\
Utah State University \\
\texttt{soheila.farokhi@usu.edu}
\and
Kelvyn Bladen \\
Utah State University \\
\texttt{kelvyn.bladen@usu.edu}
\and
Hamid Karimi \\
Utah State University \\
\texttt{hamid.karimi@usu.edu}
\and
Kevin R. Moon \\
Utah State University \\
\texttt{kevin.moon@usu.edu}
}

\date{} 

\begin{document}
\maketitle

\begin{abstract}
\end{abstract}


\section{Introduction}
\label{sec:introduction}

Graphs are essential for modeling complex relationships, analyzing networks, and offering versatile representations that capture diverse data structures~\citep{cheng2024comprehensive,chen2025data, aumon2025random, kheiri2023analysis,kerby2024learning, karimi2023analysis,correa2023manifold, moore2022temporal,javadi2024wiki, rhodes2023gaining, moon2021ensemble, boomari2020computing,duque2023diffusion, maaten2008visualizing, moon2019visualizing}. Graph Neural Networks (GNNs)~\citep{zhou2020graph}  directly process graph-structured data using the relational information encoded in a graph topology. GNNs excel in tasks such as node classification, link prediction, and graph classification across various domains such as social networks~\citep{karimi2019multi,farokhiasonam2024}, biology~\citep{li2021graph}, and education~\citep{farokhi2023enhancing,yaramala2024navigating,karimi2020online}. Their scalability, efficiency, and versatility make them invaluable for learning from large-scale graph datasets and solving a wide range of real-world problems efficiently and effectively. 
\begin{figure*}
    \centering
    \includegraphics[width=\textwidth]{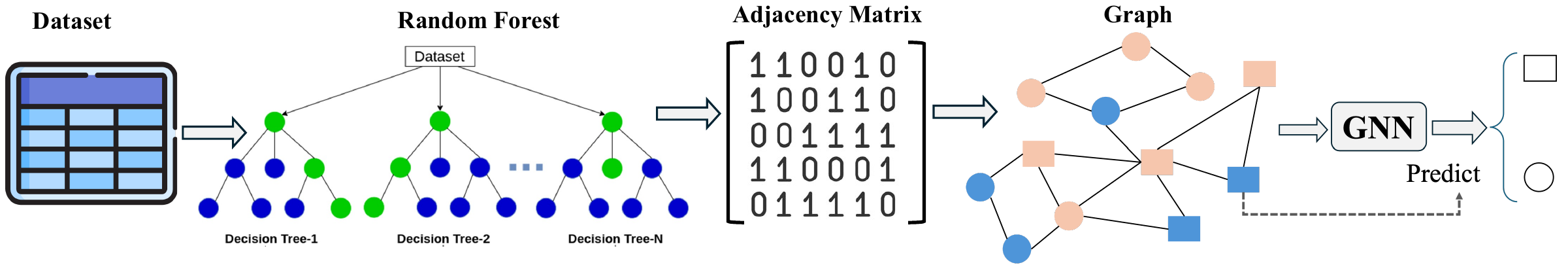}
    \vspace{-5pt}
    \caption{An overview of the proposed method (RF-GNN). A random forest is first trained on the tabular data. Pairwise proximities are extracted from the random forest and then converted to an adjacency matrix, which is used as input to a GNN.}
    \label{fig:overal-rf-gnn}
\end{figure*}

However, not all types of data possess an \textit{explicit} graph structure, particularly tabular data which is ubiquitous in the real world. 
Unlike graph-structured data, which explicitly encodes relationships between entities using nodes and edges, tabular data often lacks such explicit relational information. Instead, tabular datasets are commonly assumed to be independent and identically distributed (i.i.d.), meaning that each observation is drawn from the same underlying distribution and is independent of other observations. Nevertheless, studies have shown that breaking this assumption and creating explicit links (edges) between records -- i.e., representing the tabular data in graph form -- can be beneficial, enabling the use of GNNs to improve downstream task performance such as credit loan classification~\citep{li2023graph}. 

In this paper, we introduce RF-GNN, a  method aimed at improving the performance of traditional machine learning methods by transforming tabular data into graph-structured data, which is used as input to a GNN. Figure~\ref{fig:overal-rf-gnn} shows an overview of our approach. Specifically, we propose using  Random Forest (RF) proximities to construct a graph. Random forest proximities offer advantages over other similarity measures like Jaccard and cosine similarity, as well as kernel methods like RBF. RF proximities can handle heterogeneous data (e.g. a mixture of categorical and continuous data), capture nonlinear relationships, and perform automatic feature selection, enhancing model robustness and interpretability. Their ensemble nature provides robustness to noise and outliers, making them suitable for real-world datasets. In addition, RF proximities incorporate information about the supervised task into the graph, which is not as straightforward for other common approaches. 

Following the calculation of RF proximity values, we apply a threshold to these values to construct an \emph{adjacency matrix}, effectively capturing the relationships between samples and facilitating the generation of pseudo-graphs based on the specified threshold. This adjacency matrix is then used as input to a GNN. 

We present extensive experimental results on 36 different datasets that showcase promising outcomes. These results suggest that converting tabular data into a graph has the potential to augment the efficacy of conventional machine learning algorithms, particularly when an optimal threshold is carefully selected. 
In summary, our contributions are as follows:

\begin{itemize}
    \item We introduce a novel, general  method based on RF proximities that transforms tabular data into a graph structure for use in a GNN for supervised learning tasks.
    \item We conduct extensive experiments on 36 benchmark classification datasets encompassing a wide range of problems, dataset sizes, and features.
    \item Our method, RF-GNN, achieves up to a 0.51 improvement in the F1-score over traditional machine learning algorithms, including the original RF that was used to construct the proximities. RF-GNN consistently performs on par with or surpasses the current state of the art, including an alternative approach to applying a GNN to tabular data. This indicates that combination of an RF with a GNN results in a machine learning algorithm that is more powerful than either an RF or GNN on their own.
\end{itemize}

The remainder of this paper is organized as follows: in Section~\ref{sec:related_work}, we review related research in this area. Section~\ref{sec:background} presents the background, followed by the formal problem statements and notations in Section~\ref{sec:problem_statement}. In Section~\ref{sec:methodology}, we describe the proposed method in detail. Section~\ref{sec:experiments} is dedicated to experiments and discussions. Finally, we conclude the paper in Section~\ref{sec:conclusion} and suggest future directions.
\section{Related Work}
\label{sec:related_work}

With graph representation learning gaining significant popularity in recent years, there have been numerous efforts to represent tabular data in graph form and apply graph embedding techniques for downstream classification or regression tasks. A non-exhaustive collection of such methods, with varying degrees of code availability, is documented in the community-curated \texttt{awesome-GNN4TDL} repository \footnote{See~\url{https://github.com/Roytsai27/awesome-GNN4TDL}}. Supervised graph structure learning methods, often referred to as Deep Graph Learning (DGL), have been explored in several studies~\citep{chen2020iterative, franceschi2019learning, li2020deeprobust, zhu2021deeprobust}. These approaches use  GNNs to construct and refine the graph topology. They parameterize the adjacency matrix using various models, such as probabilistic models~\citep{franceschi2019learning, velickovic2018graph}, full parameterization models~\citep{wang2021graph}, or metric learning models~\citep{chen2020iterative, Bahare2021}. Further, they optimize both the adjacency matrix and GNN parameters through downstream task optimization. Subsequently, unsupervised DGL methods have emerged, such as the one proposed by \cite{liu2022towards}, which operates without relying on label information, thereby mitigating potential biases in learned edge distributions. However, these approaches prioritize learning improved graphical structures over directly converting tabular data into graphs. IDGL (Iterative Deep Graph Learning)~\citep{chen2020iterative} employs a weighted cosine similarity to refine graph learning iteratively, enhancing both the graph structure and node embeddings. While they improve GNN learning by refining noisy graphs and introducing graph structures to non-graphical data, direct conversion of tabular data into graph form is not their focus. 

On the other hand, some methods focus on constructing graphs by labels. \cite{rocheteautong2021} proposed linking patients with similar diagnoses in a graph to predict outcomes using an LSTM and GNN. However, its reliance on diagnosis outcomes for graph construction limits its generality. Additionally, \cite{li2024graph} listed numerous methodologies tailored to tabular datasets, some involving direct or indirect graph construction techniques like k-nearest neighbors (NN) and similarity measurement. Yet, model effectiveness crucially hinges on selecting suitable $k$ values and similarity measures. The same issue persists in~\cite{skang2021}, which used kNN to construct graphs. Moreover, most of these methods rely on unsupervised graph construction, which may be less informative than supervised alternatives~\citep{10089875}.

In summary, existing studies still lack a \textit{general} and \textit{straightforward} approach with \textit{nonrestrictive assumptions} that can directly convert tabular data into a graph and apply GNNs for downstream tasks.


\section{Background}
\label{sec:background}
In this section, we review the foundational methodologies that underpin our study, including RFs and RF proximities.

\textbf{Random Forests}~\citep{rf1,rf2} are ensemble learning methods that construct a collection of decision trees using bootstrap sampling. For each tree, a bootstrap sample is drawn with replacement from the training data. Samples that are selected in the bootstrap sample to train a given tree are referred to as \emph{in-bag} samples, while the remaining samples, which are not selected for training, are referred to as \emph{out-of-bag (OOB)} samples. On average, approximately $63.2\%$ of the training samples are in-bag for any given tree, with the remainder serving as out-of-bag observations. RFs are widely recognized as robust predictors~\citep{ho1995random}. They typically perform well ``out of the box" as they tend to require little to no tuning, especially for classification. They are adapted for both classification and regression tasks, offer straightforward parallelization, accommodate mixed variable types (continuous and categorical), remain unaffected by monotonic transformations, demonstrate resilience to outliers, scale adeptly to datasets of varying sizes, handle missing values effectively, capture non-linear interactions, and exhibit robustness to noise variables. 

\textbf{Random Forest Proximity}: Random forests naturally generate pair-wise proximity measures based on the partitioning space of their constituent decision trees. Several RF proximity measures exist. The original proximity measure was defined by Leo Breiman as the proportion of trees in which observations share the same terminal node~\citep{rfprox}. This encodes a supervised similarity measure, leveraging the optimized splitting variable values for the task at hand. Used in various applications such as data visualization~\citep{pang2006pathway,finehout2007cerebrospinal,rhodes2021random,rhodes2023gaining}, outlier detection~\citep{pang2006pathway,nesa2018outlier}, and data imputation~\citep{pantanowitz2009missing,kokla2019random,shah2014comparison}, RF proximities extend many unsupervised problems to a supervised context effectively. 

A more robust proximity metric called Random Forest-Geometry- and Accuracy-Preserving proximities (RF-GAP) was introduced in~\cite{10089875}. The proximity-weighted sum (regression) or majority vote (classification) using RF-GAP precisely replicates the out-of-bag random forest prediction, thus capturing the data geometry learned by the model. It outperforms the original random forest proximities in tasks such as data imputation, outlier detection, and visualization.






\global\long\def\x{\mathbf{x}}
\global\long\def\X{\mathcal{X}}
\global\long\def\P{\mathbf{P}}
\global\long\def\Y{\mathcal{Y}}
\global\long\def\y{\mathbf{y}}
\global\long\def\R{\mathbb{R}}
\global\long\def\T{\mathcal{T}}
\global\long\def\D{\mathcal{D}}
\section{Problem Statement}
\label{sec:problem_statement}

Let $\mathbf{\D} = \{(\x_1, y_1), (\x_2, y_2), \ldots, (\x_N, y_N)\}$ be the given tabular dataset, where $\x_i \in \R^d$ are the data points  and $y_i \in \mathbb{N}_0$ are the labels. For each dataset, we use $c \geq 2$ to indicate the number of classes. We use $\X$ to represent all data points, i.e. $\X = \{\x_1, \x_2, \ldots, \x_N\}$ and $\Y$ to represent labels, i.e. $\Y = \{y_1, y_2, \ldots, y_N\}$. Our objective is to first develop a model $f(\cdot)$ that, given a tabular dataset \( \D \), constructs a graph where data points serve as nodes and edges between these nodes indicate that the data points exhibit high similarity. We can write this as:
\begin{equation}
    A = f(\X, \Y_{train}), 
\end{equation}
where $\X$ is defined above, $\Y_{train}$ is a subset of labels $\Y$ chosen for training model $f(\cdot)$, and $A$ is the adjacency matrix for the resulting graph. Then we employ a Graph Neural Network (GNN) that, given the above adjacency matrix, node features $\X$, and target labels $\Y_{train}$ learns to predict labels of test data points $\Y_{test}$. The model's predictive function can be mathematically represented as follows:
\begin{equation}
    \hat{y_i} = \phi(A, \X, \Y_{train}), 
\end{equation}
where $\hat{y_i} \in \R^c$ is the vector that represents the estimated probability of data point $x_i$ belonging to each possible class, and $\phi$ represents any GNN model that is designed for node classification.

\section{The Proposed Method (RF-GNN)}
\label{sec:methodology}
\begin{figure*}
    \centering
    \includegraphics[width=0.9\textwidth]{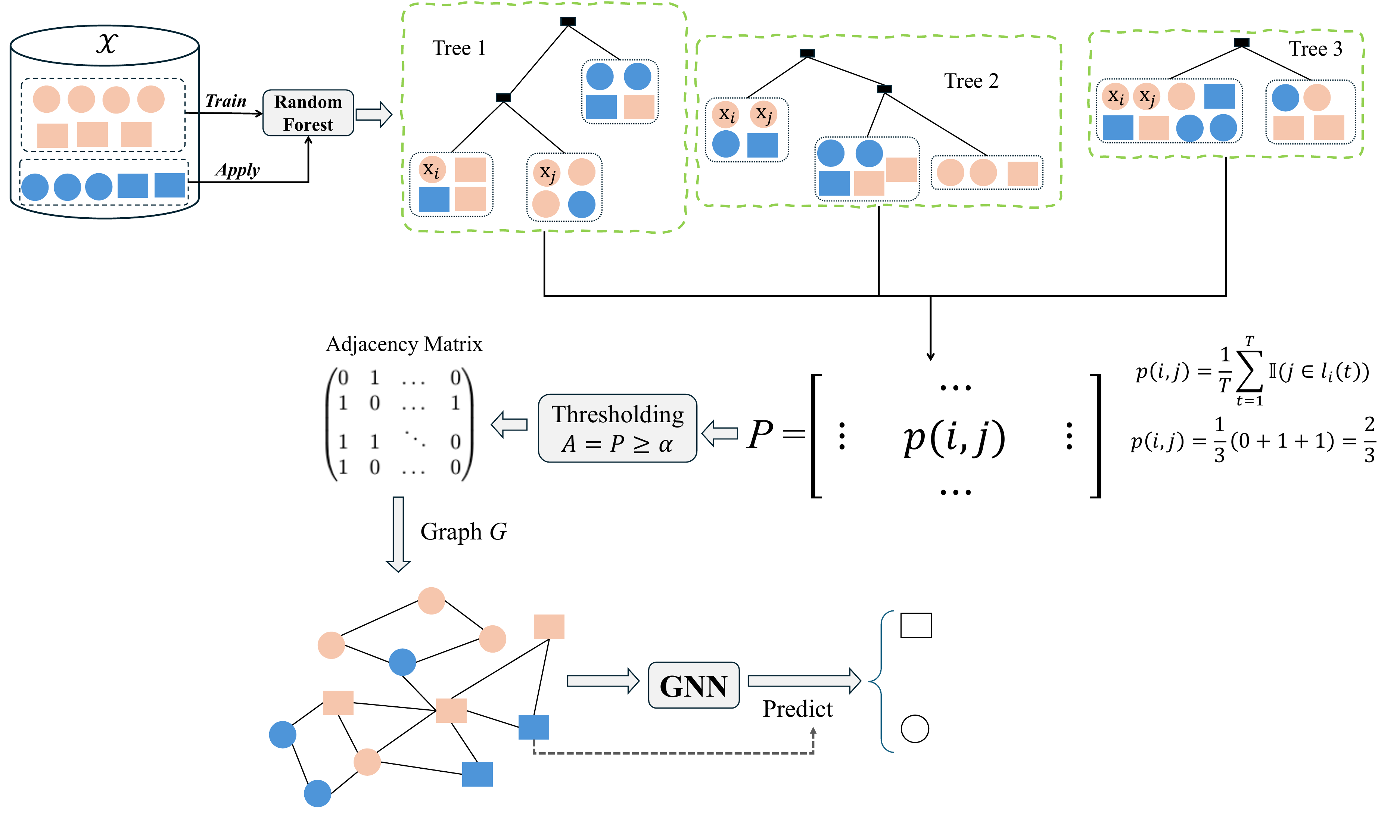}
    \vspace{-10pt}
    \caption{The workflow of our proposed method (RF-GNN), which uses a random forest to learn a graph from tabular data. The graph structure is then used as input to a GNN for final prediction. }
    \label{fig:model}
\end{figure*}

In this section, we detail RF-GNN, a novel approach that combines the robust feature extraction capabilities of random forests with the dynamic relational modeling strength of GNNs. Figure~\ref{fig:model} illustrates our proposed method. RF-GNN first transforms tabular data into a structured graph format,  described in Section~\ref{subsec:graph_const}.  In Section~\ref{subsec:gnn}, we elaborate on how the graph-structured data are processed using Graph Convolutional Networks (GCNs). 


\subsection{Graph Construction}
\label{subsec:graph_const}

To leverage graph embedding techniques, we construct a graph that captures the similarity between instances in a tabular dataset using RF proximities. The original proximity measure proposed by Breiman~\cite{rfprox} quantifies the similarity between pairs of data points based on their co-occurrence in the same terminal node (leaf) across multiple decision trees within a random forest model. High proximity indicates that data points are ``close" or ``similar" within the forest's decision-making context, providing insights into relationships in the feature space. This method is useful for tasks such as cluster identification and outlier detection~\citep{10089875} and as a foundation for other machine-learning models.

We begin by performing a 5-fold cross-validation grid search on the training set to identify an optimal Random Forest (RF) model. Specifically, we tune the number of trees (\texttt{n\_estimators} $\in \{50, 100, 200, 500, 700, 1000\}$), the minimum number of samples required to split an internal node (\texttt{min\_samples\_split} $\in \{2, 5, 10\}$), and the minimum number of samples required at a leaf node (\texttt{min\_samples\_leaf} $\in \{1, 20, 50, 80, 100, 150, 200, 300, 500\}$).

After selecting the best model, we fit it to the training data. Then, we apply the model to the whole dataset, which applies the trees in the forest to all data samples and returns leaf indices. In other words, for each data point $\x\in\X$ and for each tree in the forest, this method returns the index of the leaf that $\x$ ends up in each tree without fitting the model to the whole data. The original RF proximity between observations \(i\) and \(j\) is computed as follows:
\begin{equation}
\label{eq:rf_prox}
    p_{Or}(i,j) = \frac{1}{T} \sum_{t=1}^{T} \mathbbm{1} (j \in l_i (t)),
\end{equation}
where \(T\) represents the number of trees, \(l_i(t)\) are the indices of observations in the same leaf as \(x_i\) in tree \(t\), and \(\mathbbm{1}(\cdot)\) is the indicator function. This equation determines the proportion of trees where observations \(i\) and \(j\) share the same terminal node. For instance, in Figure~\ref{fig:model}, $p(i,j)=\frac{2}{3}$, indicating $x_i$ and $x_j$ appear in the same leaf nodes in two trees out of the three decision trees (Tree 2 and Tree 3). 

We also provide  two alternative RF proximity measures. One is called the OOB  Proximity~\citep{liaw2002classification,hastie2001tibshirani}, which is defined as
\begin{equation}
    \label{eq:oob_prox}
    p_{OOB}(i,j) = \frac{\sum_{t\in S_i} \mathbbm{1}(j\in O(t) \cap v_i(t))}{\sum_{t\in S_i} \mathbbm{1}(j\in O(t))},
\end{equation}
where $O(t)$ denotes the set of indices of the observations that are
out-of-bag in tree $t$, $S_i$ is the set of trees for which observation $i$ is out-of-bag, and $v_i(t)$ denotes the terminal node of $i$ in tree $t$. In other words, this proximity measures the proportion of trees in which observations $i$ and $j$ reside in the same terminal node, both being out-of-bag.

While random forest proximities have been used as a measure of similarity between observations, both the original and the OOB definitions exhibit drawbacks. The original definition computes proximities using all observations without considering whether they were in-bag or OOB. Since trees are grown to purity, in-bag points from different classes are necessarily separated into different terminal nodes, which artificially inflates class separation. As an example, suppose there are two Blue points ($\textcolor{blue}{x_1}, \textcolor{blue}{x_2}$) and two Red points ($\textcolor{red}{x_3}, \textcolor{red}{x_4}$). Blue--Red pairs always receive proximity zero from the trees in which they are both in-bag, exaggerating the distinction between classes even if the points are close in the feature space. 

The OOB definition attempts to correct this by using only observations that were OOB for a given tree. But this approach has two key limitations. First, it excludes in-bag data entirely, despite the fact that random forest predictions are built from in-bag votes or averages. Second, it does not account for the number of in-bag samples per leaf, unlike the weighted majority voting used in prediction. In the toy example, $\textcolor{blue}{x_1}$ and $\textcolor{blue}{x_2}$ (both Blue) never co-occur as OOB samples and therefore appear less similar than they should, while leaves with very different training sample sizes are treated equivalently. As a result, neither proximity definition faithfully captures the geometry or predictive behavior learned by the forest.

The second alternative proximity measure that we adopt in this work is the RF-GAP  proposed by~\cite{10089875}. Let $B(t)$ denote the multiset of bootstrap (in-bag) indices for tree $t$, which may contain repeated entries. We define $J_i(t)$ as the set of in-bag observations that share the same terminal node with observation $i$ in tree $t$, i.e., $J_i(t) = B(t) \cap v_i(t)$. The corresponding multiset of in-bag observations, including repetitions, is denoted by $M_i(t)$. For each $j \in J_i(t)$, let $c_j(t)$ be the multiplicity of $j$ in $B(t)$. Then, for two observations $i$ and $j$, their RF-GAP proximity is defined as
\begin{equation}
    \label{eq:rfgap_prox}
    p_{\mathrm{GAP}}(i,j) = \frac{1}{|S_i|} \sum_{t \in S_i} \frac{c_j(t) \cdot \mathbbm{1}(j \in J_i(t))}{|M_i(t)|},
\end{equation}
where $S_i$ is the collection of trees in which $i$ is out-of-bag. In words, the proximity between observations $i$ and $j$ is computed as a weighted average across trees, where the contribution of $j$ depends both on its multiplicity in the bootstrap sample and the reciprocal of the number of in-bag observations that share a terminal node with $i$. Intuitively, RF-GAP adjusts proximities so that they better reflect the geometry learned by the forest while also preserving the predictive behavior, by weighting co-occurrences according to how strongly $j$ was represented during training.

However, as RF-GAP proximities are computed using weighted sums of training points, their test-test proximity is inherently zero. To address this limitation, we adopted a Markov-diffusion–based approach. 
Specifically, let $P_{\text{test,train}}$ and 
$P_{\text{train,test}}$ denote the $n_{\text{test}} \times n_{\text{train}}$ and 
$n_{\text{train}} \times n_{\text{test}}$ cross-proximities, respectively. 
We construct the missing $n_{\text{test}} \times n_{\text{test}}$ proximity matrix as: 
\[
P_{\text{test,test}} \;=\; 
P_{\text{test,train}} \; P_{\text{train,test}}.
\]
This formulation can be interpreted as a diffusion process in which similarities between test samples are inferred indirectly via their shared relationships with the training set. Intuitively, two test points 
are considered close if they are both strongly connected to similar 
training points. This construction preserves the geometry encoded by the training proximities while extending the proximity notion to unseen test data, ensuring that the resulting graph structure remains coherent and expressive for downstream GNN tasks.

Using one of Equations~\ref{eq:rf_prox} - \ref{eq:rfgap_prox}, we calculate the proximity between all pairs of points, resulting in a proximity matrix \(P \in \mathbb{R}^{N \times N}\) where \(P_{i,j} = p(i,j)\). Using $P$ directly as the adjacency matrix results in a highly dense (or fully connected) graph, which would negatively impact the performance of GNNs and increase both storage and computational complexity. Thus we select a threshold \(\alpha\) to transform the matrix \(P\) into a binary adjacency matrix \(A \in \mathbb{R}^{\{0,1\}}\):
\begin{equation}
    A = P \geq \alpha.
\end{equation}
Thus, \(A\) can serve as the adjacency matrix for an unweighted graph with nodes as data points \(x_i\) and edges representing that the random forest proximity exceeds the threshold \(\alpha\). In Section~\ref{subsec:alpha}, we analyze the effect and distribution of this threshold.

In our experiments, none of the three proximity measures exhibited a universal advantage over the others in terms of performance. Thus the choice of which proximity measure to use to construct the graph is left as a hyperparameter.

\subsection{Graph Neural Network}
\label{subsec:gnn}
By transforming tabular data into a graph format, we redefine our initial classification problem as a node classification task. GNNs have proven highly effective in this domain by exploiting both node features and their interconnections within the graph~\citep{kipf2016semi, velickovic2017graph, hamilton2017inductive}. In particular, GCNs harness the graph's topology to aggregate and refine features from adjacent nodes~\citep{kipf2016semi}. This approach allows nodes to integrate information from their immediate and extended neighborhood, adapting the principles of convolutional neural networks to graph data~\citep{o2015introduction}. As a result, GCNs are proficient at handling various graph-related tasks, including node classification, graph classification, and link prediction.

In the GCN framework, multiple graph convolutional layers are used for neighborhood aggregation. Initially, at the 0-th layer, a node's embedding vector, or latent representation, is set to its feature vector. For instance, for node \(i\) corresponding to data point \(x_i\), we define \( h_i^{(0)} = x_i \), where \( h_i^{(0)} \) represents the embedding of node \(i\) at the 0-th layer. 

In subsequent layers \(l\), within a GCN consisting of \(L\) layers, the embeddings of nodes are updated through a weighted average of the embeddings of their neighbors, combined with the node's own previous layer's embedding, followed by the application of a non-linear activation function. This process is mathematically expressed as:
\begin{equation}
     h_i^{(l+1)} = \sigma \left( W_l \sum_{j \in N(i)} \frac{h_j^{(l)}}{|N(i)|} + B_l h_i^{(l)} \right), \forall l \in \{0, \ldots, L-1\},
\end{equation}
where \(\sigma\) denotes a non-linear function such as the Rectified Linear Unit (ReLU), and \(W_l\) and \(B_l\) are learnable parameters. \(W_l\) is the weight matrix at layer \(l\) applied to the neighbors' embeddings, \(B_l\) is the weight matrix applied to the node's own embedding from the previous layer, the neighborhood $N(i)$ of node $i$ is defined by the adjacency matrix $P$ constructed from Random-Forest-induced proximities, and \(|N(i)|\) indicates the count of these neighbors. After processing through \(L\) layers, the final embedding of node \(i\) is a \(k\)-dimensional vector given by \( h_i^{(L)} \).

After applying the GCN, the final embedding of each node or data point is used to predict its label. For this purpose, we use a two-layer Multilayer Perceptron (MLP), also known as a fully connected network. This MLP takes the final node embedding as input and predicts the probabilities of the node belonging to each class. The mathematical formulation of the MLP operations is given by:
\begin{align}
    a &= \text{ReLU}(W_{\text{in}} \cdot h_i^{(L)} + b_a), \label{eq:mlp_1} \\
    \hat{y}_i &= W_{\text{out}} \cdot a + b_o. \label{eq:mlp_2}
\end{align}

In Equation~\ref{eq:mlp_1}, \( W_{\text{in}} \) and \( b_a \) are learnable parameters. \( W_{\text{in}} \) is the weight matrix for the input layer to the hidden layer, \( b_a \) is the bias for the hidden layer, and \( a \) is the output of the hidden layer. In Equation~\ref{eq:mlp_2}, \( W_{\text{out}} \) and \( b_o \) are learned by the model. \( W_{\text{out}} \) is the weight matrix connecting the hidden layer to the output layer, \( b_o \) is the bias for the output layer, and \( \hat{y}_i \) is a \( c \)-dimensional vector representing the probability of data point \( x_i \) belonging to each class. Finally, a softmax function is applied to \( \hat{y}_i \) to determine the final label. We used a cross-entropy loss function to train the model for classification.

\section{Experiments}
\label{sec:experiments}

\subsection{Datasets}
 We use datasets available through the OpenML\footnote{\url{https://www.openml.org/}} API in our experiments. Most of our datasets belong to OpenML-CC18 benchmark~\citep{bischl2017openml}, which comprises real-world classification datasets specifically curated for benchmarking machine learning algorithms. It is widely recognized and commonly used for tabular data classification. For our analysis, we selected 36 datasets, each featuring a mix of numerical and categorical attributes. These datasets are diverse in terms of the number of instances, number of features, number of classes, and their data sources. Table~\ref{tab:dataset_stats} details the statistics of these datasets. We partitioned the datasets into training and test sets following an 80:20 split ratio.

\begin{table}[t!]
\centering
\caption{Dataset Statistics: $\mathbf{N}$ (\# of instances), $\mathbf{d}$ (\# of features), $\mathbf{d_{cat}}$ (\# of categorical features), $\mathbf{c}$ (\# of classes).}
\label{tab:dataset_stats}
\setlength{\tabcolsep}{3pt} 
\renewcommand{\arraystretch}{1.1} 

\scalebox{0.9}{
\begin{tabular}{c|c|c|c|c|c}

\textbf{ID} & \textbf{Name} & $\mathbf{N}$ & $\mathbf{d}$ & $\mathbf{d_{cat}}$ & $\mathbf{c}$\\  \hline \hline 

 902 & Sleuth Case 2002 & 147 & 6 & 4 & 2 \\
\highlightrow 1006 &  Lymphography & 148 & 18 & 15 & 2 \\
955 & Teaching Assistant Evaluation & 151 & 5 & 2 & 2 \\
\highlightrow 941 & Low Birth Weight  & 189 & 9 & 7 & 2 \\
1012 & Nations' Flags & 194 & 28 & 26 & 2 \\
\highlightrow 446 & Cushing's Syndrome & 200 & 7 & 1 & 2 \\
40710 & Heart Disease & 303 & 13 & 8 & 2 \\
\highlightrow 915 & Plasma Retinol&315 & 13 & 3 & 2 \\
1167 & ``pc1\_req" & 320 & 8 & 1 & 2 \\
\highlightrow 40663 & calendarDOW & 399 & 32 & 20 & 5 \\
475 &German Political System & 400 & 5 & 4 & 4 \\
\highlightrow 1498 & South Africa Heart Disease & 462 & 9 & 1 & 2 \\
853 & Boston Housing Data & 506 & 13 & 1 & 2 \\
\highlightrow 825 &Corrected Boston Housing Data & 506 & 20 & 3 & 2 \\
43757 & Wisconsin Breast Cancer & 569 & 30 & 0 & 2 \\
\highlightrow 40981 & Australian Credit Approval & 690 & 14 & 8 & 2 \\
43942 & Annealing & 898 & 38 & 32 & 2 \\
\highlightrow 40705 &Tokyo SGI Server Performance & 959 & 44 & 2 & 2 \\
31 & German Credit 1 & 1000 & 20 & 13 & 2 \\
\highlightrow 44098 & German Credit 2& 1000 & 20 & 13 & 2 \\
43255 & Student Performance & 1000 & 7 & 4 & 2 \\
\highlightrow 983 & Contraceptive Method Choice & 1473 & 9 & 7 & 2 \\
23 & Contraceptive Method Choice (Multi) & 1473 & 9 & 7 & 3 \\
\highlightrow 720 & Abalone Age Prediction & 4177 & 8 & 1 & 2 \\
1557 & Abalone Age Prediction (Multi) & 4177 & 8 & 1 & 3 \\
\highlightrow  40701 & Churn & 5000 & 20 & 4 & 2 \\
182 &Landsat Satellite & 6430 & 36 & 0 & 6 \\
\highlightrow  300 &Isolated Letter Speech Recognition & 7797 & 617 & 0 & 26 \\
1478 &Human Activity Recognition & 10299 & 561 & 0 & 6 \\
\highlightrow  1053 &Software Defect Prediction & 10885 & 21 & 0 & 2 \\
32 & Pen-Based Recog of Handwritten Digits& 10992 & 16 & 0 & 10 \\
\highlightrow 4534 & Phishing Websites & 11055 & 30 & 30 & 2 \\
6 & Letter Image Recognition & 20000 & 16 & 0 & 26 \\
\highlightrow  1486 & Nomao (Search engine of places) &34465 & 118 & 29 & 2 \\
1461 & Portuguese Bank Marketing&45211 & 16 & 9 & 2 \\
\highlightrow 1590 & Adult (Census Income) &48842 & 14 & 8 & 2 \\
\hline
\hline
\end{tabular}
}
\end{table}

\subsection{Baselines}
 We compare our model with the following baseline models:
\subsubsection{Random Forest (RF)} For our experiments, we employed the Random Forest implementation provided by scikit-learn\footnote{\url{https://scikit-learn.org/stable/}}.

\subsubsection{XGBoost (XGB)} XGBoost is a highly efficient and flexible gradient boosting library that enhances the speed and performance of gradient-boosted trees. It is designed for scalability and handles large-scale data effectively. We employed the XGBoost Python package\footnote{\url{https://xgboost.readthedocs.io/en/stable/python/}} in our experiments.

\subsubsection{LightGBM (LGBM)} LightGBM is a fast, distributed gradient boosting framework that uses tree-based learning algorithms. It is optimized for speed and memory efficiency, using advanced techniques like histogram-based splits. In our experiments, we used the LightGBM python package\footnote{\url{https://lightgbm.readthedocs.io/en/latest/Python-Intro.html}}.

\subsubsection{Gradient Boosting (GB)} Gradient Boosting constructs a predictive model through a sequential ensemble of weak models, typically decision trees. It optimizes arbitrary differentiable loss functions, making it versatile for various regression and classification problems. For our experiments, we used the Gradient Boosting implementation available in scikit-learn\footnote{\url{https://scikit-learn.org/stable/}}.

\subsubsection{Multi-layer Perceptron (MLP)} MLP is a type of feedforward artificial neural network with multiple layers, using backpropagation for training. It excels in complex pattern recognition by modeling non-linear relationships between inputs and outputs. For our experiments, we implement an MLP with two hidden layers using PyTorch\footnote{\url{https://pytorch.org/}}. The hidden layer dimensions are set to $[64, 128]$ and $[128, 256]$ for different model configurations. 

\subsubsection{Interaction Network Contextual Embedding (INCE)} In addition to the above ML algorithms, we compare RF-GNN with the latest relevant method from the literature. In INCE~\citep{villaizan2023graph}, both categorical and continuous features are projected individually into a common dense latent space. The resulting feature embeddings are then structured into a fully connected graph, augmented with an additional virtual node. Subsequently, a GNN is employed to model the relationships among all nodes, including the original features and the virtual node, thereby enhancing their representations. The enhanced representation of the virtual node is then input into the final classifier or regressor. For this model, we used the authors' implementation available on GitHub\footnote{\url{https://github.com/MatteoSalvatori/INCE}}.

\subsection{Experimental Results}
For each model, we employed grid search to find the best hyper-parameters. We repeated each experiment 5 times with different random seeds and recorded the mean and standard deviation of weighted F1-scores. The experiments were conducted on a system with an AMD EPYC 7513 CPU, 4 NVIDIA RTX A4000 GPUs, and 1 TB of RAM. In RF-GNN, we used PyTorch to implement the GCN component and scikit-learn for the random forest. Since RF-GNN requires a proximity threshold, for each dataset, we experimented with 51 evenly spaced proximity thresholds $\alpha$ ranging from 0 to 1 (inclusive). We then picked the best proximity threshold and model hyper-parameters using 5-fold cross-validation on the training set and applied the best model on the test set. 

\begin{table*}[p]
\setlength{\tabcolsep}{2.0pt}
\renewcommand{\arraystretch}{0.95}
\centering
\caption{Comparison of baseline models and proposed methods using weighted F1-score across datasets. Proposed methods are shown in blue; best and second-best results are highlighted in red and orange. RF-Prox, RF-GAP, and OOB denote RF-GNN variants constructed with different proximity matrices, and RF-GNN reports the best-performing variant per dataset. Standard deviations over multiple seeds are reported in Appendix~\ref{app:std}.}

\label{tab:exp_results}

\begin{tabular}{c|ccccccccccc}
\toprule
\textbf{Dataset} & \textbf{RF} & \textbf{XGB} & \textbf{GB} & \textbf{LGBM} & \textbf{MLP} & \textbf{INCE} & \textbf{{\textcolor{blue}{RF\_Prox}}} & \textbf{{\textcolor{blue}{RFGAP}}} & \textbf{\textcolor{blue}{OOB}} & \textbf{\textcolor{blue}{RF\_GNN}} \\  \hline \hline 
902 & 0.827 & 0.782 & 0.745 & 0.812 & 0.800 & \textcolor{orange}{0.833} & \textcolor{red}{0.845} & 0.813 & 0.816 & \textcolor{red}{0.845} \\
\highlightrow 1006 & 0.904 & 0.871 & 0.875 & 0.876 & 0.873 & 0.893 & \textcolor{red}{0.947} & \textcolor{orange}{0.920} & 0.901 & \textcolor{red}{0.947} \\
955 & 0.787 & 0.777 & 0.732 & 0.606 & 0.590 & 0.822 & \textcolor{orange}{0.857} & \textcolor{red}{0.869} & 0.856 & \textcolor{red}{0.869} \\
\highlightrow 941 & 0.764 & 0.693 & 0.756 & 0.709 & 0.501 & \textcolor{red}{0.812} & 0.784 & 0.783 & \textcolor{orange}{0.785} & \textcolor{orange}{0.785} \\
1012 & 0.421 & 0.407 & 0.412 & 0.426 & 0.430 & \textcolor{orange}{0.686} & 0.642 & 0.732 & \textcolor{red}{0.738} & \textcolor{red}{0.738} \\
\highlightrow 446 & 0.879 & 0.895 & 0.905 & 0.901 & 0.715 & \textcolor{red}{1} & \textcolor{red}{1} & \textcolor{orange}{0.975} & \textcolor{orange}{0.975} & \textcolor{red}{1} \\
40710 & 0.796 & 0.741 & 0.765 & 0.755 & 0.645 & \textcolor{red}{0.872} &\textcolor{orange}{0.832}  & 0.823 & 0.822 & \textcolor{orange}{0.832} \\
\highlightrow 915 & 0.556 & 0.586 & 0.589 & 0.579 & 0.447 & 0.572 & \textcolor{red}{0.729} & \textcolor{orange}{0.621} & 0.619 & \textcolor{red}{0.729} \\
1167 & 0.339 & 0.323 & 0.346 & 0.280 & 0.635 & 0.629 & 0.482 & \textcolor{red}{0.695} & \textcolor{orange}{0.685} & \textcolor{red}{0.695} \\
\highlightrow 40663 & \textcolor{red}{0.675} & 0.590 & \textcolor{orange}{0.673} & 0.612 & 0.498 & 0.633 & 0.653 & 0.623 & 0.619 & 0.653 \\
475 & 0.354 & 0.403 & 0.437 & 0.371 & 0.321 & 0.414 & 0.440 & \textcolor{red}{0.453} & \textcolor{orange}{0.450} & \textcolor{red}{0.453} \\
\highlightrow 1498 & 0.531 & 0.573 & 0.559 & 0.573 & 0.651 & 0.752 & 0.638 & \textcolor{orange}{0.781} & \textcolor{red}{0.787} & \textcolor{red}{0.787} \\
825 & 0.859 & 0.831 & 0.841 & 0.841 & 0.737 & \textcolor{orange}{0.868} & \textcolor{red}{0.901} & 0.807 & 0.809 & \textcolor{red}{0.901} \\
\highlightrow 853 & \textcolor{orange}{0.909} & 0.899 & 0.905 & 0.905 & 0.750 & 0.877 & \textcolor{red}{0.915} & 0.829 & 0.820 & \textcolor{red}{0.915} \\
43757 & 0.972 & 0.974 & 0.963 & 0.968 & 0.937 & \textcolor{red}{0.979} & 0.977 & 0.977 & \textcolor{orange}{0.978} & \textcolor{orange}{0.978} \\
\highlightrow 40981 & 0.916 & 0.911 & 0.902 & 0.910 & 0.800 & 0.930 & \textcolor{orange}{0.932} & \textcolor{red}{0.933} & 0.932 & \textcolor{orange}{0.933} \\
43942 & \textcolor{red}{1} & \textcolor{red}{1} & 0.997 & \textcolor{red}{1} & \textcolor{red}{1} & \textcolor{red}{1} & \textcolor{orange}{0.999} & 0.996 & \textcolor{red}{1} \\
\highlightrow 40705 & 0.945 & 0.941 & \textcolor{orange}{0.945} & 0.946 & 0.769 & 0.940 & \textcolor{red}{0.955} & 0.933 & 0.935 & \textcolor{red}{0.955} \\
31 & \textcolor{red}{0.832} & \textcolor{orange}{0.831} & 0.826 & 0.828 & 0.577 & 0.756 & 0.827 & 0.577 & 0.580 & 0.827 \\
\highlightrow 43255 & \textcolor{red}{0.894} & 0.874 & 0.885 & 0.883 & 0.863 & 0.791 & 0.886 & 0.888 & \textcolor{orange}{0.891} & \textcolor{orange}{0.891} \\
44098 & \textcolor{red}{0.832} & \textcolor{orange}{0.831} & 0.826 & 0.828 & 0.578 & 0.719 & 0.829 & 0.577 & 0.573 & 0.829 \\
\highlightrow 983 & 0.641 & 0.624 & 0.641 & 0.631 & 0.674 & \textcolor{red}{0.729} & 0.664 & \textcolor{orange}{0.726} & 0.725 & \textcolor{orange}{0.726} \\
23 & 0.526 & 0.530 & 0.521 & 0.517 & 0.484 & 0.484 & \textcolor{red}{0.546} & 0.530 & \textcolor{orange}{0.531} & \textcolor{red}{0.546} \\
\highlightrow 720 & 0.773 & 0.773 & 0.772 & 0.769 & 0.728 & 0.788 & \textcolor{red}{0.796} & 0.790 & \textcolor{orange}{0.792} & \textcolor{red}{0.796} \\
1557 & 0.658 & 0.656 & 0.652 & 0.653 & 0.561 & \textcolor{red}{0.672} & \textcolor{orange}{0.668} & 0.658 & 0.657 & \textcolor{orange}{0.668} \\
\highlightrow 40701 & 0.870 & 0.871 & 0.866 & 0.883 & 0.798 & \textcolor{orange}{0.885} & \textcolor{red}{0.885} & 0.874 & 0.874 & \textcolor{red}{0.885} \\
182 & 0.918 & \textcolor{orange}{0.919} & \textcolor{red}{0.924} & 0.917 & 0.829 & 0.873 & 0.917 & 0.912 & 0.910 & 0.917 \\
\highlightrow 300 & 0.938 & 0.952 & 0.948 & \textcolor{orange}{0.962} & 0.939 & 0.698 & \textcolor{red}{0.964} & 0.941 & 0.939 & \textcolor{red}{0.964} \\
1478 & 0.976 &\textcolor{orange}{0.991} & 0.990 & \textcolor{red}{0.992} & 0.944 & 0.924 & 0.982 & 0.976 & 0.977 & 0.982 \\
\highlightrow 1053 & 0.329 & 0.292 & 0.262 & 0.242 & 0.738 & \textcolor{red}{0.770} & \textcolor{orange}{0.758} & 0.734 & 0.735 & \textcolor{red}{0.758} \\
32 & 0.990 & 0.989 & 0.992 & 0.990 & 0.950 & 0.980 & \textcolor{orange}{0.993} & \textcolor{red}{0.994} & 0.994 & \textcolor{red}{0.994} \\
\highlightrow 4534 & \textcolor{red}{0.976} & 0.969 & 0.970 & 0.969 & 0.914 & 0.969 & 0.971 & 0.974 & \textcolor{orange}{0.975} & \textcolor{orange}{0.975} \\
6 & \textcolor{red}{0.966} & 0.964 & \textcolor{orange}{0.966} & 0.965 & 0.690 & 0.775 & 0.939 & 0.863 & 0.864 & 0.939 \\
\highlightrow 1486 & \textcolor{orange}{0.979} & \textcolor{red}{0.980} & 0.978 & 0.977 & 0.933 & 0.954 & 0.966 & 0.595 & 0.595 & 0.966 \\
1461 & 0.510 & 0.559 & 0.544 & 0.544 & 0.847 & \textcolor{red}{0.910} & 0.892 & 0.897 & \textcolor{orange}{0.897} & \textcolor{orange}{0.897} \\
\highlightrow 1590 & 0.683 & 0.714 & 0.719 & 0.713 & 0.731 & \textcolor{red}{0.818} & \textcolor{orange}{0.764} & 0.657 & 0.658 & \textcolor{orange}{0.764} 
\\ \hline \hline
\# 1st rank & 6 & 2 & 1 & 2 & 1 & \textcolor{orange}{10} & 11 & 5 & 3 & \textcolor{red}{18}\\
\highlightrow \# 2nd rank & 2 & \textcolor{orange}{4} & 3 & 1 & 0 & \textcolor{orange}{4} & 8 & 5 & 10 & \textcolor{red}{10} \\
Model rank & 3 & 4 & 5 & 6 & 7 & \textcolor{orange}{2} & & & & \textcolor{red}{1} \\

\bottomrule
\end{tabular}
\end{table*}

Table~\ref{tab:exp_results} and Table~\ref{tab:prox_diff} shows the results of our experiments. Based on these results, we make the following observations:






\begin{itemize}
    \item \textbf{RF-GNN} Our method demonstrates strong empirical performance, achieving the top rank on 18 out of 36 datasets and the second rank on an additional 10, leading to the best overall average rank. This highlights the effectiveness of integrating Random Forest proximities with GNNs, which successfully capture structural relationships within tabular data and thereby enhance predictive accuracy.

    \item \textbf{INCE} is the second-best model with an average rank of 2, leading in performance on 10 datasets and achieving second place on 4 datasets. Its robust performance highlights the benefits of graph embedding techniques for tabular data classification.
    
    \item The standard \textbf{Random Forest} emerges as a strong contender, holding a close third position with an average rank of 3. It consistently performs well, emphasizing its reliability as a robust method for diverse datasets.
    
    \item The superior performance of \textbf{RF-GNN} and \textbf{INCE} across the majority of the datasets (28 out of 36) illustrates their particular effectiveness in leveraging graph-based approaches to improve predictions in tabular datasets.

    \item While \textbf{RF-GNN} generally outperforms \textbf{INCE}, it shows notably higher scores on datasets such as 43255 and 6, indicating significant improvements. Conversely, \textbf{INCE} demonstrates better results than \textbf{RF-GNN} on datasets such as 1498 and 983, suggesting that the choice between these models may depend on specific dataset characteristics or the nature of the data relationships captured by each model.

    \item Within the RF-GNN framework, three proximity definitions are available as selectable parameters: the original proximity (RF\_Prox), RF-GAP (RFGAP), and the out-of-bag proximity (OOB). Overall, the original proximity achieves the highest average predictive performance across datasets (Table~3), and there are several datasets where RF\_Prox substantially outperforms RFGAP or OOB (e.g., datasets 915, 825, 853, and 31). At the same time, there are also multiple datasets in which RFGAP or OOB clearly outperform the original proximity (e.g., datasets 1012, 1167, 1498, and 983), indicating that no single proximity uniformly dominates across all settings.
    
    \item On average, RFGAP and OOB exhibit nearly identical predictive performance, with only a negligible mean difference between the two (Table~3). While both alternatives achieve slightly lower mean performance than RF\_Prox, their variance across datasets is comparable, and their relative behavior is highly consistent. Consequently, average performance alone does not meaningfully distinguish between RFGAP and OOB.
    
    \item In practice, when selecting proximities, we recommend always including the original RF proximity, as it achieves the highest average predictive performance across datasets. To limit computational cost while still allowing flexibility, one may optionally include either RFGAP or OOB as an alternative proximity. Since RFGAP and OOB exhibit nearly identical predictive accuracy, we slightly favor RFGAP due to its principled geometric construction and its consistent empirical behavior.

\end{itemize}

\begin{table}
\centering
\caption{Summary statistics of predictive performance across proximity constructions.
Top rows report mean performance differences; bottom rows report absolute performance
(mean $\pm$ standard deviation) across datasets.}
\label{tab:prox_diff}
\begin{tabular}{lcc}
\toprule
 & Mean & Std. \\
\midrule
\multicolumn{3}{l}{\textit{Average performance differences}} \\
RFGAP -- RF\_Prox & -0.029 & 0.102 \\
OOB -- RF\_Prox   & -0.030 & 0.102 \\
RFGAP -- OOB      & 0.001 & 0.005 \\
\midrule
\multicolumn{3}{l}{\textit{Absolute performance}} \\
RF\_Prox          & \textbf{0.827} & 0.151 \\
RFGAP             & 0.798 & 0.149 \\
OOB               & 0.797 & 0.149 \\
\bottomrule
\end{tabular}
\end{table}

\subsection{Comparison with other Proximity Measures}

In this section, we compare the usage of the RF-based proximities with other similarity measures in a GNN.   To evaluate the efficacy of the RF proximities compared to such measures, we opted to use cosine, Jaccard, and RFB kernel measures where for each pair of samples $x_i$ and $x_j$, they will yield a similarity scalar between 0 and 1 based on the features of these two samples. The following are mathematical formulations of these similarity measures:
\begin{equation}
    Cosine (x_i, x_j)= \frac{x_i \cdot x_j}{||x_i||\times ||x_j||},
\end{equation}
where $\cdot$ is the inner product operation, and $||\cdot||$ is the Euclidean norm operation.

\begin{equation}
    Jaccard(x_i, x_j) = \frac{\sum_{k=1}^d \min(x_{i,k}, x_{j,k})}{\sum_{k=1}^d \max(x_{i,k}, x_{j,k})},
\end{equation}
where $x_{i,k}$ is the $k$-th element in the $d$-dimensional vector $x_i$.

\begin{equation}
    RBF(x_i, x_j) = \exp\left(-\gamma \|x_i - x_j\|^2\right),
\end{equation}
where $||\cdot||$ is the Euclidean norm operation and $\gamma$ is a hyper-parameter that determines the bandwidth of the kernel. We set the value of $\gamma$ to $0.01$ in our experiments. We also standardized the RBF value to be in the range $[0,1]$.

\begin{figure}
\centering
\includegraphics[width=1\columnwidth]{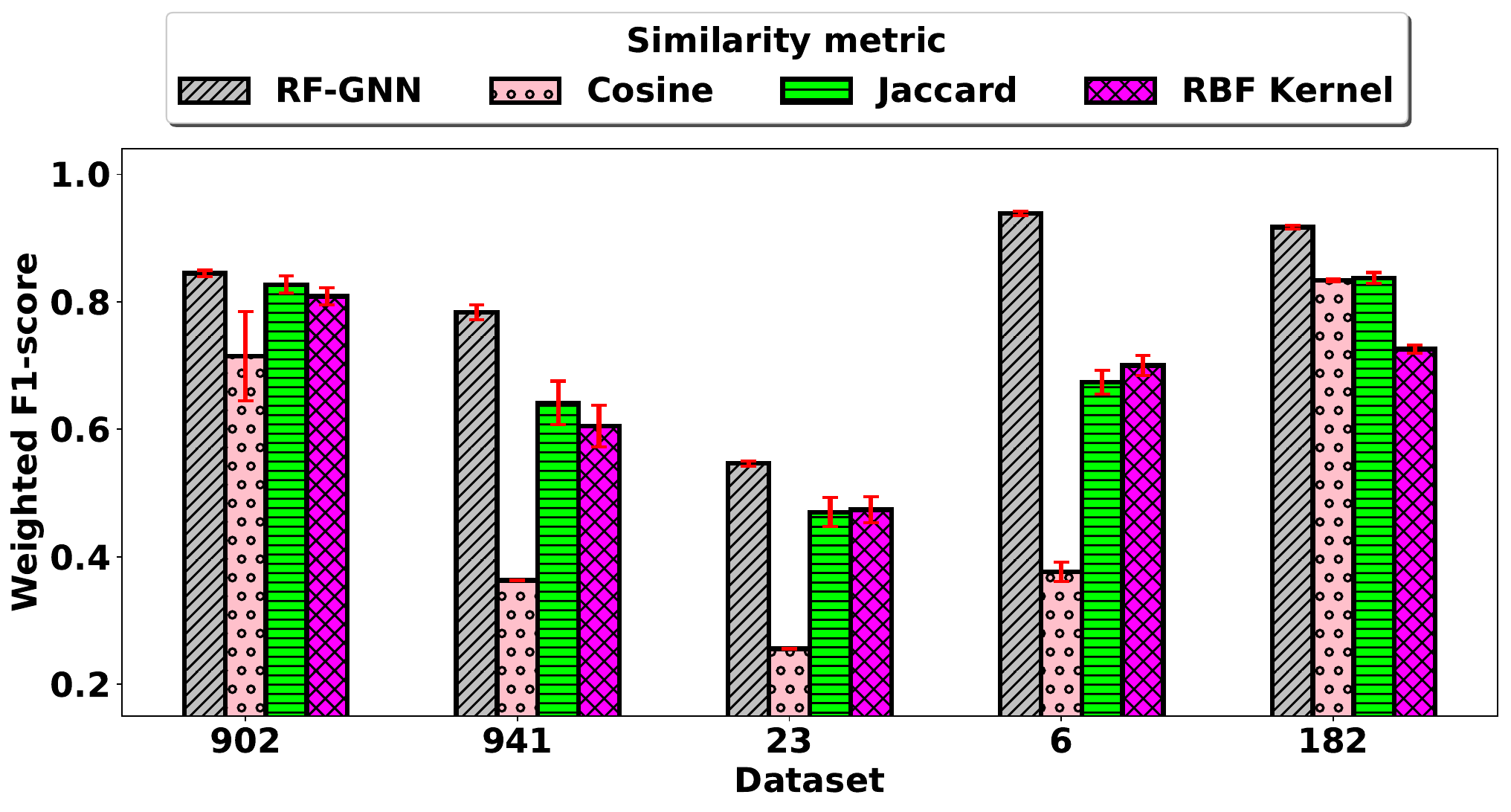}
\vspace{-20pt}
\caption{Effect of using different proximity measures on model performance on 5 different datasets in terms of weighted F1-score. The RF proximity gives the best performance.}
\label{fig:similarity}


\end{figure}

We compare the weighted F1-score of RF-GNN with GNNs trained on graphs constructed using the aforementioned similarity measures across 5 different datasets\footnote{We obtained similar results for other datasets.}. We employ the same training scheme, constructing proximity matrices for the dataset via all these similarity measures and then applying a threshold to construct graphs for GNNs. The results are shown in Figure~\ref{fig:similarity}. Each experiment was executed 5 times with different random seeds, and the mean and standard deviation of weighted F1-scores are reported. We make the following observations.
\begin{itemize}
    \item Our experimental results in Figure~\ref{fig:similarity} show that RF-GNN consistently outperforms GNNs trained on graphs constructed from other proximity measures.
    \item Unlike traditional proximity measures that solely focus on the feature space, RF proximities consider both features and labels, providing a more comprehensive understanding of sample similarity and enhancing performance in tasks where label information is informative or critical.
    \item  RF proximities, trained within the context of a random forest model, inherently prioritize the most informative features with respect to the supervised task during the learning process.
    \item RF proximities tends to be less sensitive to outliers compared to other similarity measurements due to the ensemble nature of the Random Forest algorithm and the way proximities are calculated within it. 
\end{itemize}

\subsection{Threshold Sensitivity Analysis}
\label{subsec:alpha}

\begin{figure}[htb]
\centering
\includegraphics[width=\textwidth]{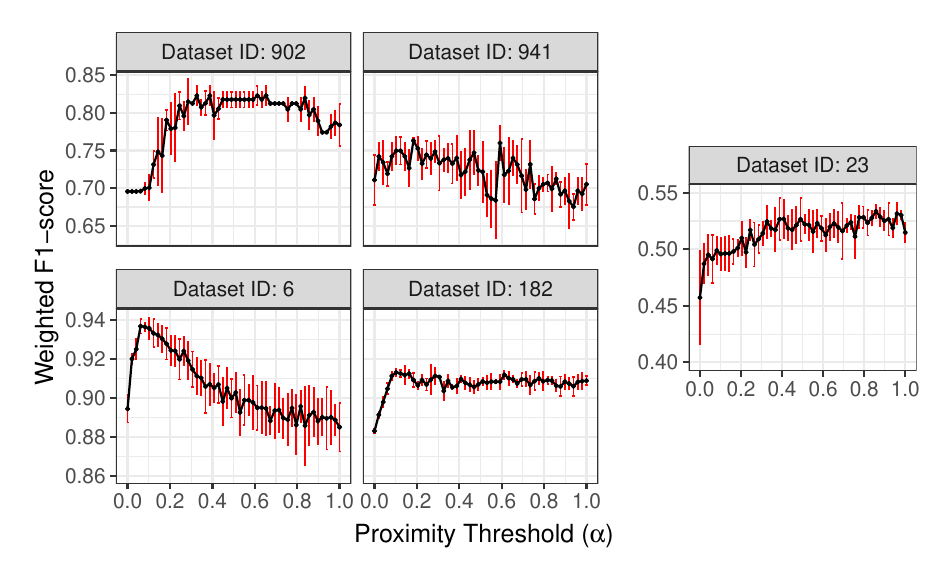}
\vspace{-0.3cm}
\caption{
Sensitivity analysis of RF-GNN performance across varying proximity thresholds $\alpha$ for five datasets (902, 941, 6, 182, and 23). 
The weighted F1-score is reported with error bars over repeated runs. 
Across most datasets, performance remains stable for threshold values in the range $\alpha \in [0.2, 0.4]$. 
Dataset 941 exhibits relatively higher variability, indicating greater sensitivity to threshold selection. 
Overall, the results demonstrate that RF-GNN is robust to the edge-threshold hyperparameter within a moderate range.
}
\label{fig:proximity_threshold}
\end{figure}


We conducted a sensitivity analysis to determine the impact of various proximity thresholds $\alpha$ on the performance of RF-GNN. This analysis was performed across five of the datasets in Table~\ref{tab:dataset_stats}, using 51 evenly spaced thresholds ranging from 0 to 1. For each threshold, experiments were repeated five times, and both the mean and standard deviation of the weighted F1-scores were recorded. The results are illustrated in Figure~\ref{fig:proximity_threshold}.

The analysis indicates that, apart from some minor variations between seeds, the performance for datasets generally stabilizes for $\alpha$ values between 0.4 and 0.6. However, dataset 941 shows fluctuating performance at certain thresholds, suggesting variable sensitivity to the proximity threshold setting.

\begin{figure}
\centering
\includegraphics[width=0.8\columnwidth]{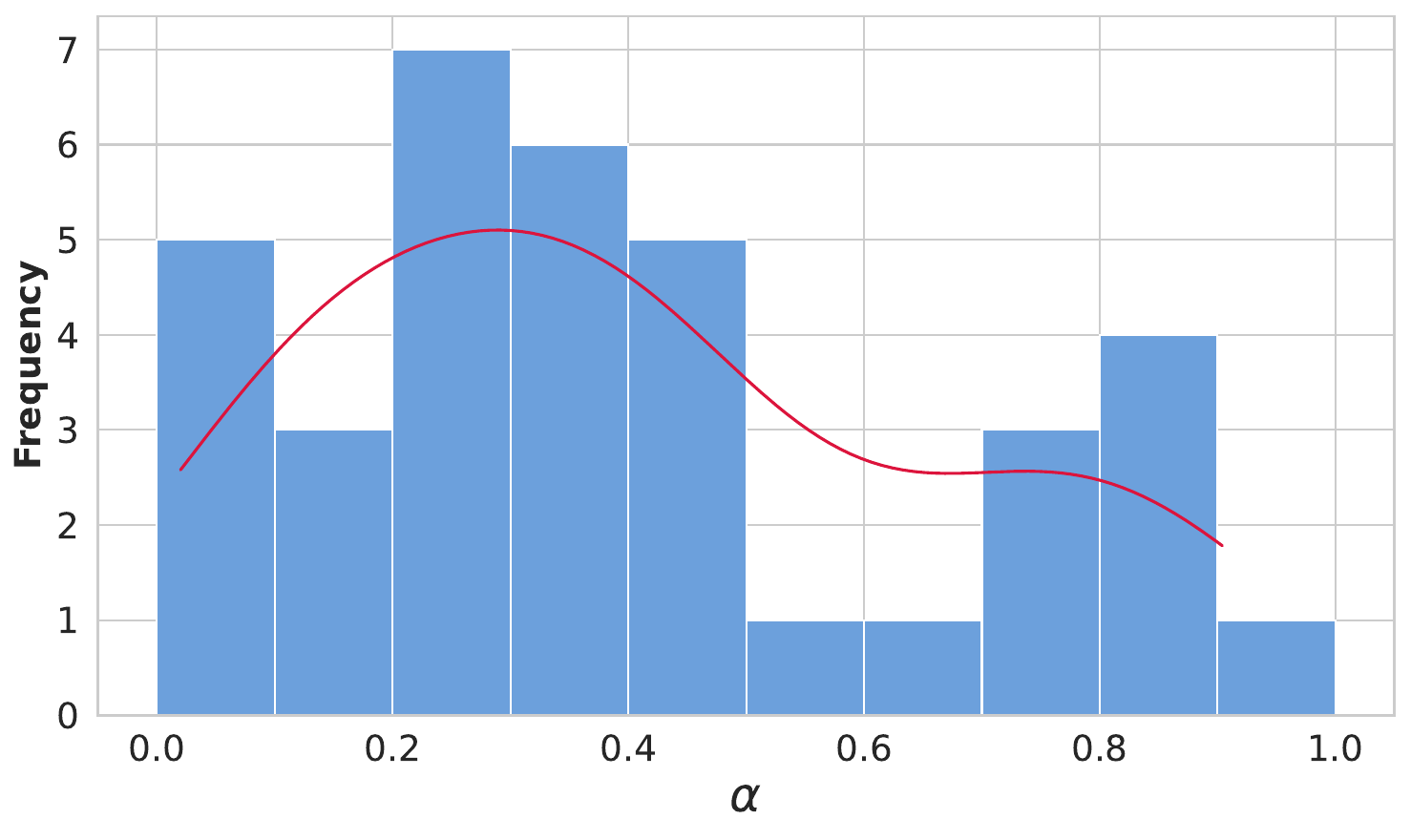}
\vspace{-10pt}
\caption{Distribution of optimal proximity thresholds $\alpha$ across 36 datasets, highlighting a concentration between 0.1 and 0.5. This range suggests that moderate graph density is optimal for GNN performance.}
\label{fig:alpha_pdf}
\end{figure}

Additionally, we examined the distribution of the optimal proximity thresholds across all 36 datasets by averaging the best-performing thresholds from each of our 5 experiments. The resulting histogram and Kernel Density Estimate (KDE) plot, shown in Figure~\ref{fig:alpha_pdf}, demonstrate that optimal $\alpha$ values typically range from 0.1 to 0.5 with a peak around 0.3. This pattern indicates that moderately dense graph structures, which strike a balance between connectivity and noise, tend to yield the best performance in GNNs.

\subsection{Dataset Attribute Analysis}
\label{subsec:dataset_attr}
While RF-GNN outperforms all other methods on average, there are a few instances where another method outperforms RF-GNN by at least a few percentage points (e.g. datasets 40710 and 1590). 
Here we investigate whether coarse dataset-level attributes help explain variations in predictive performance across methods. Specifically, we consider weighted F1-score as the dependent variable and the number of instances, number of features, number of classes, and number of categorical features as independent variables. Full regression results for all methods are reported in Appendix~\ref{app:regression_full}. Here, we focus on the most informative and statistically significant findings, which are only present when considering the difference between RF-GNN models based on the original proximities and alternative RF proximities.

\paragraph{Difference-based proximity analysis.}
To better understand when alternative proximity constructions are beneficial, we introduce \emph{difference-based} regressions that model performance gaps between RF-GNN variants. In particular, we define:
\begin{itemize}
    \item \textbf{Diff\_RFGAP$(+)$} and \textbf{Diff\_RFGAP$(-)$} as the positive and negative performance differences between RF-GNN constructed using RF-GAP proximity and the original RF proximity.
    \item \textbf{Diff\_OOB$(+)$} and \textbf{Diff\_OOB$(-)$} analogously capture the performance differences between RF-GNN using out-of-bag (OOB) proximity and the original RF proximity.
\end{itemize}
The $(+)$ models capture scenarios where the alternative proximity outperforms the original RF proximity, while the $(-)$ models capture cases where it underperforms.

\begin{table}
\centering
\caption{Summary of statistically significant relationships in difference-based proximity regressions. Positive coefficients indicate that the corresponding dataset attribute increases the magnitude of the performance difference.}
\label{tab:diff_summary}
\scriptsize
\setlength{\tabcolsep}{4pt}
\renewcommand{\arraystretch}{1.1}
\begin{tabular}{l c c c}
\toprule
Model & Attribute & Sign & $p$-value \\
\midrule
Diff\_RFGAP$(-)$ & \#Instances & $+$ & 0.0106 \\
Diff\_OOB$(-)$   & \#Instances & $+$ & 0.0094 \\
Diff\_RFGAP$(+)$ & \#Classes & $-$ & 0.0346 \\
Diff\_RFGAP$(+)$ & \#Categorical Features & $-$ & 0.0466 \\
\bottomrule
\end{tabular}
\end{table}

\paragraph{Key findings.}
The most notable and statistically significant relationships emerge in the difference-based regressions and are summarized in Table~\ref{tab:diff_summary}. First, the number of instances exhibits a positive and significant effect in both Diff\_RFGAP$(-)$ and Diff\_OOB$(-)$. This indicates that as dataset size increases, deviations from the original RF proximity become increasingly detrimental, suggesting that the original RF proximity is particularly robust for larger datasets while alternative proximity definitions may introduce unnecessary noise.

Second, Diff\_RFGAP$(+)$ reveals significant negative effects for both the number of classes and the number of categorical features. This suggests that RF-GAP is more likely to outperform the original RF proximity in datasets with fewer classes and simpler categorical structure. As class complexity and categorical diversity increase, the advantage of RF-GAP diminishes.

\paragraph{Summary.}
Overall, while coarse dataset attributes have limited explanatory power for absolute model performance, the difference-based analysis reveals systematic patterns governing when alternative proximity constructions help or hinder RF-GNN. The original RF proximity demonstrates strong robustness across dataset scales, whereas RF-GAP and OOB proximities may be advantageous in more structured, lower-complexity settings.

\section{Conclusion and Future Work}
\label{sec:conclusion}

This work introduced RF-GNN, a general framework for enhancing tabular learning by converting feature-based data into graph representations using Random Forest–derived proximities and applying Graph Neural Networks. By leveraging both RF-GAP and out-of-bag (OOB) proximity constructions, RF-GNN effectively captures higher-order relational structure that is typically inaccessible to conventional tabular models. Extensive empirical evaluation across 36 heterogeneous datasets demonstrated that RF-GNN consistently outperforms or matches strong baselines, including Random Forests, gradient boosting methods, multi-layer perceptrons, and recent representation-learning approaches. These results highlight the robustness and broad applicability of Random Forest–based graph construction as a bridge between tabular learning and graph-based inference.

Our analysis further showed that RF-derived proximities provide a particularly effective inductive bias for graph construction, and that performance is sensitive to the choice of proximity threshold. While RF-GAP and OOB proximities mitigate several known issues of standard RF proximity measures—such as label leakage and biased co-occurrence counts—the experimental results suggest that the choice of RF proximity measure remains a critical design dimension in RF-GNN.

Several promising directions for future work arise from this observation. One immediate extension is the development of more principled and adaptive thresholding criteria. Rather than relying on fixed or heuristically chosen thresholds, future work could explore data-driven or optimization-based threshold selection, potentially guided by graph sparsity constraints, spectral properties, or downstream validation performance. Dynamic thresholding strategies that evolve during training may further improve stability and generalization.

Another important avenue is refining how proximities are weighted within the constructed graph. While RF-derived similarities are powerful, they may over-amplify confident but biased tree predictions, particularly in regions of class imbalance or limited feature diversity. Alternative weighting schemes—such as uncertainty-aware reweighting, normalization based on node degree or local density, or attenuation functions that dampen extreme proximity values—could help mitigate over-exaggeration effects while preserving meaningful relational structure.

Beyond proximity refinement, future work could investigate tighter theoretical connections between Random Forest geometry and graph learning, as well as scalability improvements for very large datasets through approximate graph construction and parallelization. Exploring alternative GNN architectures, including attention-based or inductive models, may further enhance performance when combined with RF-based graphs. Finally, extending RF-GNN to settings involving temporal, sequential, or partially observed tabular data represents a natural and impactful direction.

Overall, RF-GNN establishes a flexible and interpretable pathway for integrating tree-based learning with graph neural networks, and opens multiple avenues for both theoretical development and practical deployment in complex tabular domains.

\section*{Acknowledgements}
This work was supported in part by the NSF under Grant 2212325.

\bibliographystyle{unsrtnat} 
\bibliography{References}

\appendix
\clearpage
\section{Training Variability and Randomness}
\label{app:std}

\begin{table*}[htb]
\scriptsize
\setlength{\tabcolsep}{1.5pt}
\renewcommand{\arraystretch}{0.95}
\centering
\caption{Mean $\pm$ standard deviation of weighted F1-score across multiple seeds (corresponding to Table~\ref{tab:exp_results}).}
\label{tab:mean_std_full}
\scriptsize
\resizebox{0.95\textwidth}{!}{\begin{tabular}{c|ccccccccc}
\toprule
Dataset & RF & XGB & GB & LGBM & MLP & INCE & \textcolor{blue}{RF\_Prox} & \textcolor{blue}{RFGAP} & \textcolor{blue}{OOB} \\
\midrule
902   & 0.827 $\pm$ 0.018 & 0.782 $\pm$ 0.015 & 0.745 $\pm$ 0.032 & 0.812 $\pm$ 0.000 & 0.800 $\pm$ 0.021 & 0.833 $\pm$ 0.024 & 0.845 $\pm$ 0.005 & 0.813 $\pm$ 0.006 & 0.816 $\pm$ 0.005 \\
1006  & 0.904 $\pm$ 0.024 & 0.871 $\pm$ 0.025 & 0.875 $\pm$ 0.053 & 0.876 $\pm$ 0.036 & 0.873 $\pm$ 0.013 & 0.893 $\pm$ 0.015 & 0.947 $\pm$ 0.021 & 0.920 $\pm$ 0.023 & 0.901 $\pm$ 0.022 \\
955   & 0.787 $\pm$ 0.025 & 0.777 $\pm$ 0.025 & 0.732 $\pm$ 0.041 & 0.606 $\pm$ 0.021 & 0.590 $\pm$ 0.061 & 0.822 $\pm$ 0.057 & 0.857 $\pm$ 0.000 & 0.869 $\pm$ 0.002 & 0.856 $\pm$ 0.001 \\
941   & 0.764 $\pm$ 0.031 & 0.693 $\pm$ 0.021 & 0.756 $\pm$ 0.020 & 0.709 $\pm$ 0.026 & 0.501 $\pm$ 0.075 & 0.812 $\pm$ 0.030 & 0.784 $\pm$ 0.012 & 0.783 $\pm$ 0.012 & 0.785 $\pm$ 0.012 \\
1012  & 0.421 $\pm$ 0.000 & 0.407 $\pm$ 0.025 & 0.412 $\pm$ 0.035 & 0.426 $\pm$ 0.071 & 0.430 $\pm$ 0.024 & 0.686 $\pm$ 0.019 & 0.642 $\pm$ 0.012 & 0.732 $\pm$ 0.012 & 0.738 $\pm$ 0.011 \\
446   & 0.879 $\pm$ 0.012 & 0.895 $\pm$ 0.012 & 0.905 $\pm$ 0.013 & 0.901 $\pm$ 0.015 & 0.715 $\pm$ 0.162 & 1.000 $\pm$ 0.000 & 1.000 $\pm$ 0.000 & 0.975 $\pm$ 0.000 & 0.975 $\pm$ 0.000 \\
40710 & 0.796 $\pm$ 0.009 & 0.741 $\pm$ 0.023 & 0.765 $\pm$ 0.016 & 0.755 $\pm$ 0.010 & 0.645 $\pm$ 0.036 & 0.872 $\pm$ 0.027 & 0.832 $\pm$ 0.013 & 0.823 $\pm$ 0.014 & 0.822 $\pm$ 0.014 \\
915   & 0.556 $\pm$ 0.030 & 0.586 $\pm$ 0.022 & 0.589 $\pm$ 0.014 & 0.579 $\pm$ 0.037 & 0.447 $\pm$ 0.027 & 0.572 $\pm$ 0.018 & 0.729 $\pm$ 0.003 & 0.621 $\pm$ 0.003 & 0.619 $\pm$ 0.003 \\
1167  & 0.339 $\pm$ 0.067 & 0.323 $\pm$ 0.000 & 0.346 $\pm$ 0.057 & 0.280 $\pm$ 0.027 & 0.635 $\pm$ 0.024 & 0.629 $\pm$ 0.082 & 0.482 $\pm$ 0.013 & 0.695 $\pm$ 0.014 & 0.685 $\pm$ 0.013 \\
40663 & 0.675 $\pm$ 0.007 & 0.590 $\pm$ 0.012 & 0.673 $\pm$ 0.007 & 0.612 $\pm$ 0.006 & 0.498 $\pm$ 0.015 & 0.633 $\pm$ 0.032 & 0.653 $\pm$ 0.010 & 0.623 $\pm$ 0.011 & 0.619 $\pm$ 0.011 \\
475   & 0.354 $\pm$ 0.017 & 0.403 $\pm$ 0.016 & 0.437 $\pm$ 0.025 & 0.371 $\pm$ 0.027 & 0.321 $\pm$ 0.028 & 0.414 $\pm$ 0.178 & 0.440 $\pm$ 0.018 & 0.453 $\pm$ 0.019 & 0.450 $\pm$ 0.019 \\
1498  & 0.531 $\pm$ 0.030 & 0.573 $\pm$ 0.019 & 0.559 $\pm$ 0.030 & 0.573 $\pm$ 0.025 & 0.651 $\pm$ 0.033 & 0.752 $\pm$ 0.015 & 0.638 $\pm$ 0.028 & 0.781 $\pm$ 0.029 & 0.787 $\pm$ 0.029 \\
825   & 0.859 $\pm$ 0.006 & 0.831 $\pm$ 0.005 & 0.841 $\pm$ 0.013 & 0.841 $\pm$ 0.008 & 0.737 $\pm$ 0.025 & 0.868 $\pm$ 0.018 & 0.901 $\pm$ 0.008 & 0.807 $\pm$ 0.009 & 0.809 $\pm$ 0.009 \\
853   & 0.909 $\pm$ 0.012 & 0.899 $\pm$ 0.004 & 0.905 $\pm$ 0.015 & 0.905 $\pm$ 0.004 & 0.750 $\pm$ 0.015 & 0.877 $\pm$ 0.018 & 0.915 $\pm$ 0.010 & 0.829 $\pm$ 0.010 & 0.820 $\pm$ 0.010 \\
43757 & 0.972 $\pm$ 0.000 & 0.974 $\pm$ 0.005 & 0.963 $\pm$ 0.006 & 0.968 $\pm$ 0.006 & 0.937 $\pm$ 0.003 & 0.979 $\pm$ 0.010 & 0.977 $\pm$ 0.003 & 0.977 $\pm$ 0.004 & 0.978 $\pm$ 0.004 \\
40981 & 0.916 $\pm$ 0.006 & 0.911 $\pm$ 0.010 & 0.902 $\pm$ 0.007 & 0.910 $\pm$ 0.005 & 0.800 $\pm$ 0.023 & 0.930 $\pm$ 0.007 & 0.932 $\pm$ 0.009 & 0.933 $\pm$ 0.010 & 0.932 $\pm$ 0.009 \\
43942 & 1.000 $\pm$ 0.000 & 1.000 $\pm$ 0.000 & 0.997 $\pm$ 0.004 & 1.000 $\pm$ 0.000 & 1.000 $\pm$ 0.000 & 1.000 $\pm$ 0.000 & 0.999 $\pm$ 0.000 & 0.996 $\pm$ 0.001 & 1.000 $\pm$ 0.000 \\
40705 & 0.945 $\pm$ 0.003 & 0.941 $\pm$ 0.004 & 0.945 $\pm$ 0.007 & 0.946 $\pm$ 0.004 & 0.769 $\pm$ 0.011 & 0.940 $\pm$ 0.005 & 0.955 $\pm$ 0.002 & 0.933 $\pm$ 0.002 & 0.935 $\pm$ 0.002 \\
31    & 0.832 $\pm$ 0.013 & 0.831 $\pm$ 0.008 & 0.826 $\pm$ 0.010 & 0.828 $\pm$ 0.010 & 0.577 $\pm$ 0.000 & 0.756 $\pm$ 0.022 & 0.827 $\pm$ 0.003 & 0.577 $\pm$ 0.004 & 0.580 $\pm$ 0.003 \\
43255 & 0.894 $\pm$ 0.004 & 0.874 $\pm$ 0.005 & 0.885 $\pm$ 0.008 & 0.883 $\pm$ 0.004 & 0.863 $\pm$ 0.009 & 0.791 $\pm$ 0.244 & 0.886 $\pm$ 0.007 & 0.888 $\pm$ 0.007 & 0.891 $\pm$ 0.007 \\
44098 & 0.832 $\pm$ 0.013 & 0.831 $\pm$ 0.008 & 0.826 $\pm$ 0.010 & 0.828 $\pm$ 0.010 & 0.578 $\pm$ 0.003 & 0.719 $\pm$ 0.080 & 0.829 $\pm$ 0.003 & 0.577 $\pm$ 0.004 & 0.573 $\pm$ 0.004 \\
983   & 0.641 $\pm$ 0.011 & 0.624 $\pm$ 0.004 & 0.641 $\pm$ 0.018 & 0.631 $\pm$ 0.010 & 0.674 $\pm$ 0.021 & 0.729 $\pm$ 0.008 & 0.664 $\pm$ 0.009 & 0.726 $\pm$ 0.009 & 0.725 $\pm$ 0.009 \\
23    & 0.526 $\pm$ 0.011 & 0.530 $\pm$ 0.013 & 0.521 $\pm$ 0.010 & 0.517 $\pm$ 0.007 & 0.484 $\pm$ 0.012 & 0.484 $\pm$ 0.128 & 0.546 $\pm$ 0.004 & 0.530 $\pm$ 0.004 & 0.531 $\pm$ 0.004 \\
720   & 0.773 $\pm$ 0.002 & 0.773 $\pm$ 0.002 & 0.772 $\pm$ 0.004 & 0.769 $\pm$ 0.004 & 0.728 $\pm$ 0.002 & 0.788 $\pm$ 0.006 & 0.796 $\pm$ 0.002 & 0.790 $\pm$ 0.002 & 0.792 $\pm$ 0.002 \\
1557  & 0.658 $\pm$ 0.004 & 0.656 $\pm$ 0.004 & 0.652 $\pm$ 0.010 & 0.653 $\pm$ 0.009 & 0.561 $\pm$ 0.013 & 0.672 $\pm$ 0.004 & 0.668 $\pm$ 0.002 & 0.658 $\pm$ 0.003 & 0.657 $\pm$ 0.003 \\
40701 & 0.870 $\pm$ 0.006 & 0.871 $\pm$ 0.004 & 0.866 $\pm$ 0.006 & 0.883 $\pm$ 0.005 & 0.798 $\pm$ 0.004 & 0.885 $\pm$ 0.053 & 0.885 $\pm$ 0.002 & 0.874 $\pm$ 0.002 & 0.874 $\pm$ 0.002 \\
182   & 0.918 $\pm$ 0.002 & 0.919 $\pm$ 0.003 & 0.924 $\pm$ 0.003 & 0.917 $\pm$ 0.000 & 0.829 $\pm$ 0.003 & 0.873 $\pm$ 0.008 & 0.917 $\pm$ 0.003 & 0.912 $\pm$ 0.003 & 0.910 $\pm$ 0.003 \\
300   & 0.938 $\pm$ 0.001 & 0.952 $\pm$ 0.002 & 0.948 $\pm$ 0.000 & 0.962 $\pm$ 0.000 & 0.939 $\pm$ 0.002 & 0.698 $\pm$ 0.046 & 0.964 $\pm$ 0.002 & 0.941 $\pm$ 0.002 & 0.939 $\pm$ 0.002 \\
1478  & 0.976 $\pm$ 0.003 & 0.991 $\pm$ 0.000 & 0.990 $\pm$ 0.000 & 0.992 $\pm$ 0.001 & 0.944 $\pm$ 0.001 & 0.924 $\pm$ 0.074 & 0.982 $\pm$ 0.001 & 0.976 $\pm$ 0.001 & 0.977 $\pm$ 0.001 \\
1053  & 0.329 $\pm$ 0.009 & 0.292 $\pm$ 0.000 & 0.262 $\pm$ 0.003 & 0.242 $\pm$ 0.002 & 0.738 $\pm$ 0.007 & 0.770 $\pm$ 0.002 & 0.758 $\pm$ 0.003 & 0.734 $\pm$ 0.004 & 0.735 $\pm$ 0.003 \\
32    & 0.990 $\pm$ 0.001 & 0.989 $\pm$ 0.000 & 0.992 $\pm$ 0.000 & 0.990 $\pm$ 0.000 & 0.950 $\pm$ 0.008 & 0.980 $\pm$ 0.006 & 0.993 $\pm$ 0.000 & 0.994 $\pm$ 0.000 & 0.994 $\pm$ 0.000 \\
4534  & 0.976 $\pm$ 0.001 & 0.969 $\pm$ 0.000 & 0.970 $\pm$ 0.001 & 0.969 $\pm$ 0.001 & 0.914 $\pm$ 0.003 & 0.969 $\pm$ 0.000 & 0.971 $\pm$ 0.001 & 0.974 $\pm$ 0.001 & 0.975 $\pm$ 0.001 \\
6     & 0.966 $\pm$ 0.000 & 0.964 $\pm$ 0.000 & 0.966 $\pm$ 0.000 & 0.965 $\pm$ 0.002 & 0.690 $\pm$ 0.006 & 0.775 $\pm$ 0.041 & 0.939 $\pm$ 0.004 & 0.863 $\pm$ 0.004 & 0.864 $\pm$ 0.004 \\
1486  & 0.979 $\pm$ 0.000 & 0.980 $\pm$ 0.000 & 0.978 $\pm$ 0.000 & 0.977 $\pm$ 0.001 & 0.933 $\pm$ 0.001 & 0.954 $\pm$ 0.004 & 0.966 $\pm$ 0.001 & 0.595 $\pm$ 0.001 & 0.595 $\pm$ 0.001 \\
1461  & 0.510 $\pm$ 0.003 & 0.559 $\pm$ 0.000 & 0.544 $\pm$ 0.000 & 0.544 $\pm$ 0.006 & 0.847 $\pm$ 0.010 & 0.910 $\pm$ 0.001 & 0.892 $\pm$ 0.002 & 0.897 $\pm$ 0.002 & 0.897 $\pm$ 0.002 \\
1590  & 0.683 $\pm$ 0.002 & 0.714 $\pm$ 0.000 & 0.719 $\pm$ 0.000 & 0.713 $\pm$ 0.002 & 0.731 $\pm$ 0.008 & 0.818 $\pm$ 0.090 & 0.764 $\pm$ 0.012 & 0.657 $\pm$ 0.012 & 0.658 $\pm$ 0.012 \\
\bottomrule
\end{tabular}}
\end{table*}

\clearpage
\section{Dataset Attribute Regression Analysis Table}
\label{app:regression_full}
\begin{table*}[htb]
\scriptsize
\setlength{\tabcolsep}{1.5pt}
\renewcommand{\arraystretch}{0.95}
\centering
\caption{Full regression results relating dataset attributes to weighted F1-score across all methods.}
\label{tab:regression}
\scalebox{0.8}{
\begin{tabular}{c|cccccc}
\toprule
Method & Dataset Attribute             & Coefficient & Std. Error & z-value   & $P > |z|$   & 95\% Confidence Interval \\
\hline
 & const                 & 0.7673      & 0.2318     & 3.3100    & 0.00093  & [0.3130, 1.2216]          \\
 & \#Instances            & -0.000008531 & 0.00001263 & -0.6750   & 0.49944  & [-0.0000332, 0.0000162]   \\
RF & \#Features             & 0.001563    & 0.001428   & 1.0950    & 0.27369  & [-0.00124, 0.00436]       \\
 & \#Classes              & 0.04961     & 0.03456    & 1.4360    & 0.15111  & [-0.0181, 0.1173]         \\
 & \#Categorical Features & 0.02790     & 0.01654    & 1.6870    & 0.09161  & [-0.0045, 0.0603]         \\
\cline{2-7}
  & Mean Squared Error (MSE): 0.0325 & & &  R-squared (R$^2$): 0.145 & & \\
\hline
 & const                 & 0.7321    & 0.2306    & 3.174    & 0.0015    & [0.281, 1.183]          \\
 & \#Instances            & -0.000004584    & 0.00001276    & -0.359    & 0.7195    & [-0.0000296, 0.0000204]     \\
XGB & \#Features             & 0.001942    & 0.001443    & 1.346    & 0.1784    & [-0.000887, 0.004771]     \\
 & \#Classes              & 0.04671    & 0.03463    & 1.349    & 0.177    & [-0.0212, 0.1146]     \\
 & \#Categorical Features & 0.02459    & 0.01625    & 1.514    & 0.1301    & [-0.0073, 0.0564]     \\ \cline{2-7}
  & Mean Squared Error (MSE): 0.0357 & & &  R-squared (R$^2$): 0.144 & & \\
\hline
  & const                 & 0.7346    & 0.2291    & 3.206    & 0.00134    & [0.2856, 1.1836]          \\
  & \#Instances            & -0.000005715    & 0.00001252    & -0.456    & 0.64815    & [-0.0000303, 0.0000188]    \\
GB  & \#Features             & 0.001909    & 0.001438    & 1.327    & 0.18440    & [-0.000908, 0.004726]     \\
  & \#Classes              & 0.05078    & 0.03505    & 1.449    & 0.14738    & [-0.0179, 0.1195]     \\
  & \#Categorical Features & 0.02561    & 0.01623    & 1.579    & 0.11444    & [-0.00621, 0.05743]     \\ \cline{2-7}
  & Mean Squared Error (MSE): 0.0334 & & &  R-squared (R$^2$): 0.16 & & \\
\hline
 & const                 & 0.6813      & 0.2352     & 2.8970    & 0.00377  & [0.2213, 1.1413]          \\
 & \#Instances            & -0.000005339 & 0.00001291 & -0.4140   & 0.67915  & [-0.0000306, 0.0000200]   \\
LGBM & \#Features             & 0.001981    & 0.001454   & 1.3620    & 0.17304  & [-0.000868, 0.004830]     \\
 & \#Classes              & 0.04884     & 0.03515    & 1.3890    & 0.16473  & [-0.0201, 0.1177]         \\
 & \#Categorical Features & 0.02631     & 0.01665    & 1.5800    & 0.11412  & [-0.0063, 0.0589]         \\   \cline{2-7}
  & Mean Squared Error (MSE): 0.0334 & & &  R-squared (R$^2$): 0.158 & & \\
\hline
  & const                 & 0.6088      & 0.1887     & 3.2270    & 0.00125  & [0.2389, 0.9787]          \\
 & \#Instances            & 0.00001286  & 0.00001183 & 1.0870    & 0.27717  & [-0.0000103, 0.0000361]   \\
MLP & \#Features             & 0.002433    & 0.001290   & 1.8850    & \textcolor{orange}{0.05938}  & [-0.000095, 0.004961]     \\
 & \#Classes              & -0.0008393  & 0.02801    & -0.0300   & 0.97609  & [-0.0557, 0.0540]         \\
 & \#Categorical Features & 0.01814     & 0.01312    & 1.3830    & 0.16679  & [-0.0076, 0.0438]         \\     \cline{2-7}
  & Mean Squared Error (MSE): 0.0565 & & &  R-squared (R$^2$): 0.188 & & \\ \hline
 & const                 & 1.2930      & 0.2073     & 6.2380    & 0.0000  & [0.8867, 1.6993]          \\
 & \#Instances            & 0.000009709 & 0.00001296 & 0.7490    & 0.45400         & [-0.0000157, 0.0000351]   \\
INCE & \#Features             & 0.0005253   & 0.001108   & 0.4740    & 0.63500         & [-0.00164, 0.00270]       \\
 & \#Classes              & -0.02171    & 0.02681    & -0.8100   & 0.41800         & [-0.0743, 0.0309]         \\
 & \#Categorical Features & 0.01196     & 0.01403    & 0.8520    & 0.39400         & [-0.0156, 0.0395]          \\ \cline{2-7}
  & Mean Squared Error (MSE): 0.0283 & & &  R-squared (R$^2$): 0.072 & & \\ \hline
 & const                 & 1.3540      & 0.2067     & 6.5520    & 0.0000 & [0.9490, 1.7590]          \\
 & \#Instances            & 0.000001545 & 0.00001196 & 0.1290    & 0.89700         & [-0.0000219, 0.0000250]   \\
RF-GNN & \#Features             & 0.001563    & 0.001317   & 1.1860    & 0.23500         & [-0.00102, 0.00414]       \\
 & \#Classes              & 0.02059     & 0.03007    & 0.6850    & 0.49300         & [-0.0384, 0.0795]         \\
 & \#Categorical Features & 0.01414     & 0.01408    & 1.0050    & 0.31500         & [-0.0135, 0.0417]         \\ \cline{2-7}
 & Mean Squared Error (MSE): 0.0269	 & & &  R-squared (R$^2$): 0.109 & & 
 \\ \hline
 & const                 & -2.7400     & 0.2669     & -10.2660  & 0.0000   & [-3.2630, -2.2170]        \\
 & \#Instances            & 0.00002498  & 0.000009776 & 2.5560    & \textcolor{red}{0.0106}   & [0.0000059, 0.0000441]    \\
Diff\_RFGAP (-) & \#Features             & -0.0004165  & 0.001076   & -0.3870   & 0.6988   & [-0.00252, 0.00170]      \\
 & \#Classes              & 0.002557    & 0.02676    & 0.0960    & 0.9239   & [-0.0499, 0.0550]         \\
 & \#Categorical Features & 0.02231     & 0.01423    & 1.5680    & 0.1169   & [-0.0056, 0.0502]          \\ \cline{2-7}
 & Mean Squared Error (MSE): tbd	 & & &  R-squared (R$^2$): 0.316 & &
  \\ \hline
  & const                 & -2.2640     & 0.3768     & -6.0090   & 0.0000   & [-3.0030, -1.5250]        \\
 & \#Instances            & -0.00002341 & 0.00001457 & -1.6070   & 0.1080   & [-0.0000520, 0.0000052]   \\
Diff\_RFGAP (+) & \#Features             & 0.1267      & 0.06872    & 1.8440    & \textcolor{orange}{0.0652}   & [-0.0080, 0.2614]        \\
 & \#Classes              & -0.2272     & 0.1075     & -2.1130   & \textcolor{red}{0.0346}   & [-0.4380, -0.0165]        \\
 & \#Categorical Features & -0.1179     & 0.05922    & -1.9900   & \textcolor{red}{0.0466}   & [-0.2340, -0.0018]        \\ \cline{2-7}
 & Mean Squared Error (MSE): tbd	 & & &  R-squared (R$^2$): 0.495 & &
   \\ \hline
 & const                 & -2.7990     & 0.2588     & -10.8150  & 0.0000   & [-3.3060, -2.2920]        \\
 & \#Instances            & 0.00002506  & 0.000009647 & 2.5970    & \textcolor{red}{0.00939}  & [0.0000062, 0.0000440]    \\
Diff\_OOB (-) & \#Features             & -0.0004195  & 0.001071   & -0.3920   & 0.69514  & [-0.00256, 0.00172]      \\
 & \#Classes              & 0.004917    & 0.02638    & 0.1860    & 0.85213  & [-0.0468, 0.0566]         \\
 & \#Categorical Features & 0.02561     & 0.01388    & 1.8450    & 0.06507  & [-0.0016, 0.0528]         \\
\cline{2-7}
 & Mean Squared Error (MSE): tbd	 & & &  R-squared (R$^2$): 0.334 & &
    \\ \hline
 & const                 & -1.8640     & 0.4179     & -4.4600   & 0.0000   & [-2.6820, -1.0460]        \\
 & \#Instances            & -0.00001199 & 0.00001548 & -0.7740   & 0.43900  & [-0.0000423, 0.0000183]   \\
Diff\_OOB (+) & \#Features             & -0.01096    & 0.02047    & -0.5350   & 0.59200  & [-0.0510, 0.0301]        \\
 & \#Classes              & -0.07728    & 0.09567    & -0.8080   & 0.41900  & [-0.2650, 0.1100]         \\
 & \#Categorical Features & -0.0003383  & 0.02233    & -0.0150   & 0.98800  & [-0.0441, 0.0434]         \\
\cline{2-7}
 & Mean Squared Error (MSE): tbd	 & & &  R-squared (R$^2$): 0.213 & &
 \\  \bottomrule
\end{tabular}
}
\end{table*}

Here we provide the full results of the regression analysis of model performance. We have several observations:
    \begin{itemize}
    \item \textbf{Overall Model Fit:}
    Several models exhibit non-negligible $R^2$ values, indicating that dataset-level attributes explain a meaningful portion of the variability in predictive performance. In particular, proximity-based graph constructions show stronger explanatory power than standard baselines.

    \item \textbf{Baseline Models:}
    For conventional machine learning models, the $R^2$ values remain relatively low, suggesting that coarse dataset attributes such as dataset size, feature dimensionality, and class count provide limited explanatory power for performance variation.

    \item \textbf{RF-GNN Models:}
    RF-GNN and its variants exhibit positive $R^2$ values, indicating that performance variations are more systematically related to dataset characteristics when proximity-based graph structure is incorporated.

    \item \textbf{Difference-Based Proximity Variants:}
    The difference-based proximity constructions (Diff\_RF-GAP and Diff\_OOB) achieve the highest $R^2$ values among all models. This suggests that contrastive proximity information captures dataset-dependent structure that is relevant to downstream performance.

    \item \textbf{\#Instances Variable:}
    The number of instances shows statistically significant coefficients in several difference-based models, indicating that larger datasets tend to benefit more from proximity-difference graph constructions.

    \item \textbf{\#Features and \#Classes Variables:}
    The number of features and the number of classes exhibit mixed effects across models. While their influence is limited for baseline methods, these variables become significant in certain proximity-based settings, reflecting interactions between dataset complexity and graph construction.

    \item \textbf{\#Categorical Features Variable:}
    The number of categorical features remains weakly significant or insignificant across most models, suggesting that categorical composition alone is not a dominant factor in determining performance.

    \item \textbf{Intercept Terms:}
    The intercepts are consistently significant across models, reflecting baseline performance levels that are not fully explained by the included dataset attributes.

    \item \textbf{Implications:}
    Taken together, these results indicate that proximity-based graph methods, particularly difference-based constructions, exhibit more interpretable and systematic relationships with dataset characteristics than non-graph baselines.
\end{itemize}

Overall, the regression analysis indicates that coarse dataset-level attributes explain only a limited portion of the variability in classification performance for standard machine learning models. In contrast, RF-GNN exhibits more stable performance across datasets with varying characteristics, suggesting that its effectiveness is not strongly driven by simple factors such as dataset size, feature dimensionality, class count, or the proportion of categorical features. These results indicate that RF-GNN leverages task-adaptive structure beyond coarse dataset statistics, supporting its robustness across diverse tabular datasets.

\end{document}